\documentclass[11pt]{article}
\usepackage{amsmath,amssymb,amsfonts,amstext}
\usepackage{tabularx}
\usepackage{hyperref}
\usepackage{epsf}
\usepackage{graphicx}
\usepackage{amsfonts}
\usepackage{subfigure}
\usepackage[usenames]{color}

\newcommand{\ncd}{\newcommand}
\ncd{\mrm}    {\mathrm}
\ncd{\beq} {\begin{equation}}
\ncd{\eeq} {\end{equation}}
\ncd{\nn}{\nonumber}

\ncd{\rred}{\color[rgb]{1.0,0.1,0.5}}      
\ncd{\ggreen}{\color[rgb]{1.0,0.1,0.5}}
\ncd{\oorange}{\color[rgb]{1,.65,.25}}
\ncd{\bblue}{\color[rgb]{1,1,.1}}
\ncd{\ggrey}{\color[rgb]{0.8,0.8,0.8}}
\ncd{\wwhite}{\color[rgb]{1,1,1}}

\definecolor{Blue}{rgb}{0,0.08,0.45}
\definecolor{Magenta}{cmyk}{0.1,0.8,0,0.1}
\definecolor{Orange}{rgb}{1,0.5,0}
\begin{document}

\title{Testing Hybrid Natural Inflation with BICEP2} 
\author{ Mariana Carrillo--Gonz\'alez$^{a}$, Gabriel Germ\'an$^{a}$\footnote{Corresponding author: gabriel@fis.unam.mx}\, ,
Alfredo Herrera--Aguilar$^{b,c}$, \\
Juan Carlos Hidalgo$^{a}$, Roberto A.  Sussman$^{d}$\\
\\
{\normalsize \textit{$^a$Instituto de Ciencias F\'isicas,} }
{\normalsize \textit{Universidad Nacional Aut\'onoma de M\'exico,}}\\
{\normalsize \textit{Apdo. Postal 48-3, 62251 Cuernavaca, Morelos, Mexico.}}\\
\\
{\normalsize \textit{$^b$Departamento de F\'{\i}sica,} }
{\normalsize \textit{Universidad Aut\'onoma Metropolitana Iztapalapa,}}\\
{\normalsize \textit{San Rafael Atlixco 186, CP 09340, M\'exico D. F., Mexico.}}\\
\\
{\normalsize \textit{$^c$Instituto de F\'{\i}sica y Matem\'{a}ticas,}}
{\normalsize \textit{Universidad Michoacana de San Nicol\'as de Hidalgo,}}\\
{\normalsize \textit{Edificio C--3, Ciudad Universitaria, C.P. 58040, Morelia, Michoac\'{a}n, Mexico.}}\\
\\
{\normalsize \textit{$^d$Instituto de Ciencias Nucleares}}, 
{\normalsize \textit{Universidad Nacional Aut\'onoma de M\'exico, }}\\
{\normalsize \textit{Apdo. Postal 70-543, 04510 M\'exico D. F., Mexico.}}\\
\\
}
\date{}
\maketitle

\begin{abstract}
We analyse \textit{Hybrid Natural Inflation} in view of the recent
results for the tensor index reported by BICEP2.  
We find that it predicts a large running of the scalar spectrum which
is potentially detectable by large scale structure through measurements of clustering of galaxies 
in combination with CMB data and by $21$  $\rm{cm} $ forest observations. 
The running of the running is also relatively large
becoming close to $10^{-2}.$ Along the way, we find general  
consistency relations at which observables are subject if the
slow-roll approximation is imposed. Failure to satisfy these equations
by the values obtained for the observables in surveys would be a
failure of the slow-roll approximation itself.

\end{abstract}

\section{Introduction}  \label{Isec}

     There is a considerable interest in the recent results reported by
BICEP2 \cite{BICEP2}. If confirmed they would give an
  important boost to the community working in inflationary cosmology
since a non-vanishing tensor index $r=0.2^{+0.07}_{-0.05}$  
($r=0.16^{+0.06}_{-0.05}$ when foreground subtraction based on dust models has been carried out \cite{BICEP2}), is a very
distinctive characterisation of inflation
\cite{Kehagias,Freese,Chenga,Choudhury,Ibanez,Bonvin,Chengb}.

In particular, from this data the
scale of inflation can be inferred to lie somewhere between
$2.03\times 10^{16} \, \mathrm{GeV}$ and $2.36\times 10^{16}\,\mathrm{GeV}$, which points at
new physics at or beyond the $GUT$ scale. 

Here we study a model of inflation
\cite{Guth:1981,Linde:1982,Albrecht:1982}, based on a Goldstone
(shift) symmetry and in a hybrid mechanism: \textit{Hybrid Natural
  Inflation} \cite{Graham:2010}. The original inflationary model based
on an anomalous Abelian symmetry is the \textit{Natural Inflation}
model of Freese et
al. \cite{Freese:1990rb,Freese:1993,Freese:2008if}. In natural
inflation, quantum corrections generate a small mass term for the
aspiring Goldstone mode. A potential problem for the model is that, to
produce sufficient inflation, the scale of symmetry breaking $f$ must
be greater than the Planck scale in which case large
quantum gravity corrections could appear.

Hybrid inflation scenarios \cite{Linde:1994}, where the evolution of
more than one field is important, have become common within
the inflationary paradigm. Thus, \textit{Hybrid Natural Inflation}
\cite{Graham:2010}, in which a second field is responsible for
terminating inflation, is a well motivated inflationary model.
While this model was originally formulated to realise low scales of
inflation \cite{German:2001tz}, hybrid natural inflation finds applications beyond this
purpose.   
In hybrid natural inflation the inflaton is a pseudo-Goldstone boson with a
small mass due to the rupture of the symmetry at the quantum level or
by explicit breaking. Since the end of inflation is triggered by
a second field, the number of e-folds of
inflation $ N_ {\chi }$ depends on the auxiliary field $\chi $ allowing
for a viable model, with values of $f$ at or below the Planck scale.

The goal of this paper is to constrain the free parameters of 
hybrid natural inflation in light of the data from BICEP2. Along the
way, we find general consistency relations at which observables are
subject if the  
slow-roll approximation is imposed. We find, in particular, that
hybrid natural inflation can sustain a large scalar running index. The
presence of a large running could be a possible resolution to the
apparent tension between high values of $r$ and previous indirect
limits based on temperature measurements \cite{Ade:2013uln}. A large
running would introduce a scale into the scalar power spectrum to
suppress power on large angular scales \cite{miranda2014}. As
discussed in \cite{adshead2011} the running is potentially detectable
by large scale structure through measurements of clustering of high-redshift galaxy surveys 
in combination with CMB data. High-redshift may also be probed with the $21  \rm{cm} $ forest signal which tracks 
the density field. This could be a powerful probe to observe small-scale power spectrum (SSPS) at $k \ge 10$ Mpc$^{-1}$.  
Mass function of collapsed gas in starless minihalos can be very sensitive to the SSPS, thus providing a system 
for observing effects on the scalar running index. In \cite{Shimabukuro:2014ava} an analysis is presented 
where halo mass functions and abundance of 21 cm absorbers are shown for several assumed combinations of the 
spectral index and running.

Refs. \cite{Abazajian:2013vfg} and \cite{Abazajian:2013oma} contain discussions on a ground-based Stage IV CMB experiment, CMB-S4 with 
$\mathcal{O} (500,000)$ detectors by 2020. It is expected that observations with this polarization experiment will probe large angular 
scales corresponding to low-multipole spectra $1< l<100$  and
unambiguously detect tensor modes with large $r$. These next-order measurements will presumably  detect deviations in the power-law spectrum, parametrized as
\begin{equation}
n_{\mathrm{s}}(k) =n_{\mathrm{s}}(k_0) + \frac{d n_{\mathrm{s}}}{d \ln k} \ln\left(\frac{k}{k_0}\right)+\cdot \cdot \cdot .  \label{Shape}\\ 
\end{equation}
A detection of a non-zero running $\frac{d n_{\mathrm{s}}}{d \ln k}$, through measurements of the E-mode damping tail, could provide information about the inflationary potential or point to models other than inflation. 

Our presentation is organised as follows: in Section~\ref{Slowsec} we
determine general constraint equations 
among the observables based on the slow-roll paradigm. These
constraints make no use of the specific form of the potential. Failure
to satisfy these equations by the values obtained for the observables
in surveys would be a failure of the slow-roll approximation itself. 
Section~\ref{HNIsec} presents a discussion of natural inflation in light of the
BICEP2 data. Following this line of presentation 
we then generalise results for hybrid natural inflation where $f$
is not restricted by the number of e-folds. Finally we summarise our results 
and conclude in Section~\ref{Csec}.

\section{Slow-roll parameters, observables and model independent
  results} \label{Slowsec}

One can set constraints on the parameters of a given inflationary potential by imposing the requirement of acceptable
inflation in terms of slow-roll parameters. This then leads to the
determination of observables produced by that
potential. The usual slow-roll parameters \cite{Liddle:2000cg} which
involve the potential and its derivatives are 
\begin{equation}
\epsilon \equiv \frac{M^{2}}{2}\left( \frac{V^{\prime }}{V }\right) ^{2},\quad
\eta \equiv M^{2}\frac{V^{\prime \prime }}{V}, \quad
\xi_2 \equiv M^{4}\frac{V^{\prime }V^{\prime \prime \prime }}{V^{2}},\quad
\xi_3 \equiv M^{6}\frac{V^{\prime 2 }V^{\prime \prime \prime \prime }}{V^{3}},
\label{Slowparameters}
\end{equation}%
where primes denote derivatives with respect to $\phi$. $M$ is the
reduced Planck mass $M=2.44\times 10^{18} \,\mathrm{GeV}$ and we set
$M=1$ in what follows. In the slow-roll approximation the observables
are given in terms of the usual slow-roll parameters
\cite{Liddle:2000cg} as follows 

\begin{eqnarray}
n_{\mathrm{t}} &=&-2\epsilon =-\frac{r}{8} , \label{Slowr} \\
n_{\mathrm{s}} &=&1+2\eta -6\epsilon ,  \label{Slowns} \\
n_{\mathrm{tk}} &=&\frac{d n_{\mathrm{t}}}{d \ln k}=4\epsilon\left( \eta -2\epsilon\right), \label{Slowntk} \\
n_{\mathrm{sk}} &=&\frac{d n_{\mathrm{s}}}{d \ln k}=16\epsilon \eta -24\epsilon ^{2}-2\xi_2, \label{Slownsk} \\
n_{\mathrm{skk}} &=&\frac{d^{2} n_{\mathrm{s}}}{d \ln k^{2}}=-192\epsilon ^{3}+192\epsilon ^{2}\eta-
32\epsilon \eta^{2} -24\epsilon\xi_2 +2\eta\xi_2 +2\xi_3, \label{Slownskk} \\
\delta _{\mathrm{H}}^{2}(k) &=&\frac{1}{150\pi ^{2}}\frac{\Lambda^4}{%
\epsilon _{\mathrm{H}}},
\label{Slowdeltah} 
\end{eqnarray}
where $n_{\mathrm{t}}$ is the tensor spectral index, $r$ is the usual
tensor index or the ratio of tensor to scalar perturbations, $n_{\mathrm{tk}}$
the running of the tensor index, $n_{\mathrm{s}}$ the scalar spectral
index, $n_{\mathrm{sk}}$ its running, $n_{\mathrm{skk}}$ the running
of the running, in a self-explanatory notation. The density
perturbation at wave number $k$ is $\delta_{\mathrm{H}}^{2}(k)$ and
$\Lambda$ is the scale of inflation with $\Lambda \equiv V_{\rm
  H}^{1/4}$. Note that all these quantities are described in 
terms of the inflaton scale $\phi_{\mathrm{H}}$, at which the perturbations are 
produced, some $50-60$ e-folds before the end of inflation. 
Defining the quantity $\delta _{\mathrm{ns}}$ by $\delta
_{\mathrm{ns}}\equiv 1-n_{s}$ we note that Eq.~(\ref{Slowntk}) can be
written as a constraint equation among the observables  
\begin{equation}
n_{\mathrm{tk}} = \frac{r}{64}\left(r-8\delta_{\mathrm{ns}}\right).  \label{Slowntk2}\\ 
\end{equation}
More constraint equations can be written as above but they have little chance of being falsified in the near future 
\cite{Mariana:2014b}. 
Note that Eq.~(\ref{Slowntk2}) is a model independent constraint on the observables and should be satisfied by 
\textit{any} model 
of inflation based on the slow-roll paradigm. 
Failure to satisfy this equation by the values obtained for the
observables in surveys would indicate a departure from the slow-roll
approximation itself.  As a simple numerical example we see that the
reported values $\delta _{\mathrm{H}}=1.87 \times 10^{-5}$,
$\delta _{\mathrm{ns}}\equiv 1-n_{\mathrm{s}}=1-0.96=0.04$, and, 
$0.15 < r < 0.27$  yield $-4\times 10^{-4} < n_{\mathrm{tk}} <
-2.1\times 10^{-4}$. The values also set the
scale of inflation in the range $2.03 \times 10^{16} \mathrm{GeV} <
\Lambda <  2.36 \times 10^{16} \mathrm{GeV}.$  
We stress that these values for $n_{\mathrm{tk}}$ and $\Lambda $
are \textit{model independent}. They only depend on the  values given
to $\delta _{\mathrm{H}}$, $n_{\mathrm{s}}$, $r $ \textit{and} the
validity of the slow-roll approximation.  

Thus, according to the slow-roll approximation for single field inflation we are predicting that
$n_{\mathrm{tk}}$ will take values within the range $-4\times 10^{-4}
< n_{\mathrm{tk}} < -2.1\times 10^{-4}$. 

\noindent On the other hand, the equation for the running Eq.~(\ref{Slownsk}) can be written as
\begin{equation}
n_{\mathrm{sk}}= \frac{1}{32}\left(3r^{2}-16r\delta _{\mathrm{ns}}-64 \xi_2 \right) .
\label{Slownsk2}
\end{equation}
Thus, a determination of $n_{\mathrm{sk}}$ would be equivalent to a
determination of the parameter $\xi_2$. In hybrid natural inflation 
as well as in natural inflation,
$n_{\mathrm{sk}}$ is given by 
\begin{equation}
n_{\mathrm{sk}}= \frac{r}{32}\left(3r-16\delta _{\mathrm{ns}}+\frac{8}{f^{2}} \right) ,
\label{Slownsk3}
\end{equation}%
where $f$ is the scale of the (Goldstone) symmetry breaking. A determination of $n_{\mathrm{sk}}$ would be a determination 
of $f$ and, thus, a possible indication of new physics at or beyond
the $GUT$ scale.

\section{Testing Hybrid Natural Inflation} \label{HNIsec}

We begin with a brief discussion of natural inflation which is then generalised to hybrid natural inflation. 
Natural inflation is a single scalar field model, where the end of
inflation is determined by the steepening of the potential until $\epsilon=1$. Thus, the inflaton value at the end of
inflation, $\phi_e$, is precisely determined and, up to 
an uncertainty about the number of intermediate e-folds of
inflation, so is the inflaton value of the observable scales
$\phi_{\mathrm{H}}$. As a result natural inflation has only two free
parameters: the scale of 
inflation  $\Lambda \equiv V_{\mathrm{H}}^{1/4} $ and $f$. The natural inflation potential is given by 
\begin{equation}
V\left(\phi\right)= V_0\left( 1+\cos \left( \frac{\phi }{f}\right)
\right) \equiv V_0\left( 1 + c_\phi \right)  ,
\label{NIV}
\end{equation}%
where we conveniently have defined $c_{\mathrm{\phi}}\equiv \cos \left(
\frac{\phi }{f}\right)$. From (\ref{Slowns}) we find that the spectral index
is always less than one: $n_{\mathrm{s}}
=1-\frac{1}{4}\left(r+(\frac{2}{f})^2\right)$. Thus, $f$ is  
determined if we know $n_{\mathrm{s}} $ and $r$ and is given by
\begin{equation}
f=\frac{2}{\sqrt{4\delta _{\mathrm{ns}}-r}}.
\label{NIf1}
\end{equation}
From the reality condition of $f$, we find a bound on $r$ 
\begin{equation}
r < 4\delta _{\mathrm{ns}}\approx 0.16,
\label{NIr2}
\end{equation} 
which is just within the range reported by BICEP2.  As an example we calculate $f$ from Eq.~(\ref{NIf1}) in the case 
when $\delta _{\mathrm{ns}}\equiv 1-n_{\mathrm{s}}=1-0.96=0.04$ and $r=0.15$. 
For these values one gets $f \approx 20$ and
\begin{equation}
N\approx 50, \quad \Lambda \approx 2 \times 10^{16}\, \mathrm{GeV}, \quad n_{\mathrm{tk}}\approx -4\times 10^{-4}, \quad n_{\mathrm{sk}}\approx -8\times 10^{-4},  \quad n_{\mathrm{skk}}\approx -2\times 10^{-5 }. 
\label{NIr3}
\end{equation}%

\noindent We see that natural inflation is marginally able to accommodate
the data reported by BICEP2. The price to pay is a very large symmetry
breaking scale $f$, with $f=20$.  In Table \ref{table1} we compare these values with other
models and with results for the reported case $r=0.16^{+0.06}_{-0.05}$ when foreground subtraction based on dust models has been carried out \cite{BICEP2}.

We now investigate whether hybrid natural inflation is able to lower the value of $f$ while, at
the same time, keeping reasonable values for the observables.  
As discussed in \cite{Graham:2010}, the terms of the (hybrid) inflaton
potential relevant when density perturbations are being produced have a simple universal form corresponding to 
the slow-roll of a single inflaton field $\phi $:%
\begin{equation}
V\simeq V_0\left( 1+a\cos \left( \frac{\phi }{f}\right) \right) =
V_0\left( 1+a c_\phi \right) .
\label{HNIV}
\end{equation}%
Natural inflation corresponds to the case $a=1$ while hybrid natural inflation demands $a < 1$
\cite{German:2001tz}. The slow-roll parameters for this model are given by 
\begin{eqnarray}
\epsilon &=&\frac{1}{2}\left(\frac{M}{f}\right)^2 a^{2}\frac{1-c_{\mathrm{\phi}} ^{2}}{\left( 1+a\, c_{\mathrm{\phi}} \right)^2},
\label{HNIeps}%
\\
\eta &=&-\left( \frac{M}{f}\right)^2 a\, \frac{c_{\mathrm{\phi}}}{1+a\, c_{\mathrm{\phi}}}=
\left( \frac{M}{f}\right)^2\frac{a^{2}}{\left(1-a^{2}\right)}\left(1 \pm \frac{1}{a}\sqrt{1-\frac{2\left(1-a^{2}\right)f^{2}}{a^{2}}\epsilon}\, \right), 
\label{HNIeta}
\\
\xi_2 &=&-\left( \frac{M}{f}\right)^4 a^{2}\frac{1-c_{\mathrm{\phi}} ^{2}}{\left( 1+a\, c_{\mathrm{\phi}} \right)^2}=-2\left( \frac{M}{f}\right)^2\epsilon, \label{HNIxi2}\\
\xi_3 &=& \left( \frac{M}{f}\right)^6 a^{3}\frac{1-c_{\mathrm{\phi}} ^{2}}{\left(
  1+a\, c_{\mathrm{\phi}}\right)^3} c_{\mathrm{\phi}}=-2\left( \frac{M}{f}\right)^2\epsilon\eta. \label{HNIxi3}
\end{eqnarray}%
From Eqs.~(\ref{Slowns}) and (\ref{HNIeta}) we get an expression for $f$
\begin{equation}
f=\frac{4a}{  \left( r\left(1-3 a^{2}\right)+8a^{2}\delta _{\mathrm{ns}} +\sqrt{4 a^{2}\left(4\delta _{\mathrm{ns}}-r\right)^{2}+r^{2}\left(1-a^{2}\right)} \right)^{1/2}  }.
\label{HNIf}
\end{equation}
It is easy to check that the reality condition for $f$ is always satisfied provided $a < 1$.
When $r \leq 4\delta _{\mathrm{ns}}$,  Eq.~(\ref{HNIf}) reduces to the
natural inflation formula Eq.~(\ref{NIf1}) in the limit $a\rightarrow
1$. However Eq.~(\ref{HNIf}) is more general, valid for $r \geq
4\delta _{\mathrm{ns}}$, including all relevant values of $r$ reported by BICEP2. 

Apart from Eq.~(\ref{HNIeta}), Eqs.~(\ref{HNIeps}) to (\ref{HNIxi3}) look the same as in the natural inflation case with $a=1$. 
From Eqs.~(\ref{Slowr}), (\ref{Slowns}), (\ref{HNIxi2}) and (\ref{HNIxi3}), one can see that the expression for 
$n_{\mathrm{sk}}$ and $n_{\mathrm{skk}}$ can be respectively written as
\begin{eqnarray}
n_{\mathrm{sk}} &=&\frac{r}{32}\left(3r-16\delta _{\mathrm{ns}}+\frac{8}{f^2}\right), 
\label{HNInsk} 
\\
n_{\mathrm{skk}} &=&\frac{r}{128 }\left(3r^{2}+\frac{12 }{f^{2}}r-\left(2\delta _{\mathrm{ns}}-
\frac{1}{f^{2}}\right)32\delta _{\mathrm{ns}}\right) . 
\label{HNInskk} 
\end{eqnarray}
When $f$ is of order $\mathcal{O} (1)$, the running can simply be written as
\begin{equation}
n_{\mathrm{sk}} \approx \frac{r}{4 f^{2}} \, ,
\label{HNInskapprox}
\end{equation}
thus, $n_{\mathrm{sk}}$ behaves like a simple scaled function of the
tensor index.  From Eq.~(\ref{HNInsk}) we get
\begin{equation}
f=\left(\frac{{8 r} } {{32 n_{\mathrm{sk}}-r\left(3 r- 16\delta _{\mathrm{ns}} \right)  } }  \right)^{1/2}  .
\label{HNIf1}
\end{equation}
One can easily find bounds for  $n_{\mathrm{sk}}$, these are
\begin{eqnarray}
n_{\mathrm{sk}} &\geq & -9 \times 10^{-4} \quad  \mathrm{for} \quad r \geq 0.15\, , \label{nskpos}%
\\
n_{\mathrm{sk}} &> & 0 \quad  \mathrm{for} \quad r > \frac{16}{3} \delta _{\mathrm{ns}} \approx 0.21\, .\label{nskneg}%
\end{eqnarray}%
While natural inflation can only accommodate small negative values of $n_{\mathrm{sk}}$, hybrid natural 
inflation allows also for positive values of $n_{\mathrm{sk}}$. In Table
\ref{table1} we show some of the observables using values for $r$ reported by BICEP2. Note
that in hybrid natural inflation $f$ is any positive number restricted
only by the lower value of the scale of inflation $\Lambda $. From the  
expression for the running,  Eq.~(\ref{HNInsk}), one can see that small
values of $f$ would imply unacceptably large $n_{\mathrm{sk}}$ while
large $f$ could give negligible contribution to the running. The lack of a theory of quantum
gravity makes the contemplation of a scale $f$ above the Planck scale
a mere speculation. 
\begin{table}[htbp]
  \centering
    \begin{tabular}{|c|c|c|c|c|c|c|c|c|c|}
    \hline
         & $f$ & $a$ & $n _{\mathrm{s}}$   & $r$ & $n _{\mathrm{sk}}$  & $n _{\mathrm{skk}}$ & $n _{\mathrm{tk}}$ & $\Lambda\, (\mathrm{GeV})$ & $N$ \\
 \hline
    \textit{NI} & $8.9$ & $-$ & $0.96$ & $0.11$ & $-7.2\times 10^{-4}$  & $-2.9\times 10^{-5}$ & $-3.6\times 10^{-4}$ & $1.88\times 10^{16}$  & $51$   \\
         \hline
    \textit{NI} & $20$ & $-$ & $0.96$ & $0.15$ & $-8.0\times 10^{-4}$  & $-3.2\times 10^{-5}$ & $-4.0\times 10^{-4}$ & $2.04\times 10^{16}$  & $50$   \\
    \hline
\textit{NI} & $\infty$ & $-$ & $0.96$ & $0.16$ & $-$  & $-$ & $-$ & $-$  & $-$   \\
    \hline
    \textit{HNI} & $1$ & $0.117$ & $0.96$ & $0.11$ & $2.6\times 10^{-2}$  & $2.2\times 10^{-3}$ & $-3.6\times 10^{-4}$ & $1.88\times 10^{16}$  & $N_{\chi}$   \\
    \hline
    \textit{HNI} & $1$ & $0.136$ & $0.96$ & $0.15$ & $3.7\times 10^{-2}$  & $3.6\times 10^{-3}$ & $-4.0\times 10^{-4}$ & $2.04\times 10^{16}$  & $N_{\chi}$   \\
    \hline
    \textit{HNI} & $1$ & $0.140$ & $0.96$ & $0.16$ & $3.9\times 10^{-2}$  & $4.0\times 10^{-3}$ & $-4.0\times 10^{-4}$ & $2.07\times 10^{16}$  & $N_{\chi}$   \\
    \hline
    \textit{HNI} & $1$ & $0.156$ & $0.96$ & $0.20$ & $5.0\times 10^{-2}$  & $5.8\times 10^{-3}$ & $-3.8\times 10^{-4}$ & $2.19\times 10^{16}$  & $N_{\chi}$    \\
    \hline
    \textit{HNI} & $1$ & $0.181$ & $0.96$ & $0.27$ & $6.9\times 10^{-2}$  & $9.8\times 10^{-3}$ & $-2.1\times 10^{-4}$ & $2.36\times 10^{16}$  & $N_{\chi}$   \\
    \hline
    \end{tabular}%
  \caption{\small Numerical values of observables derived for
    the {natural inflation}  and {hybrid natural inflation}  models
    specified by Eqs.~(\ref{NIV}) and (\ref{HNIV}) respectively.
    While in natural inflation there is a constraint for $r <
    4\delta_{\mathrm{ns}}\approx 0.16 $ (see Eqs.~(\ref{NIr2})),
    hybrid natural inflation is able to accommodate all values of $r$
    reported by BICEP2. The quantity $N_{\chi}$ in hybrid natural
    inflation corresponds to the number of e-folds and its value is
    controlled by a waterfall field $\chi$ to provide the required
    amount of inflation. The first and fourth rows contain results for
    the reported case \cite{BICEP2} 
    $r=0.16^{+0.06}_{-0.05}$ with foreground subtraction based on dust
    models. The case when $r=4\delta _{\mathrm{ns}}\approx 0.16$ is
    shown in the third and sixth rows, this value corresponds to a minimum
    of $n _{\mathrm{tk}}$ according to Eq.~(\ref{Slowntk2}). Note also
    the double-value character of $n _{\mathrm{tk}}$.} 
  \label{table1}%
\end{table}%

\begin{figure}[!ht]
\includegraphics[scale=0.32]{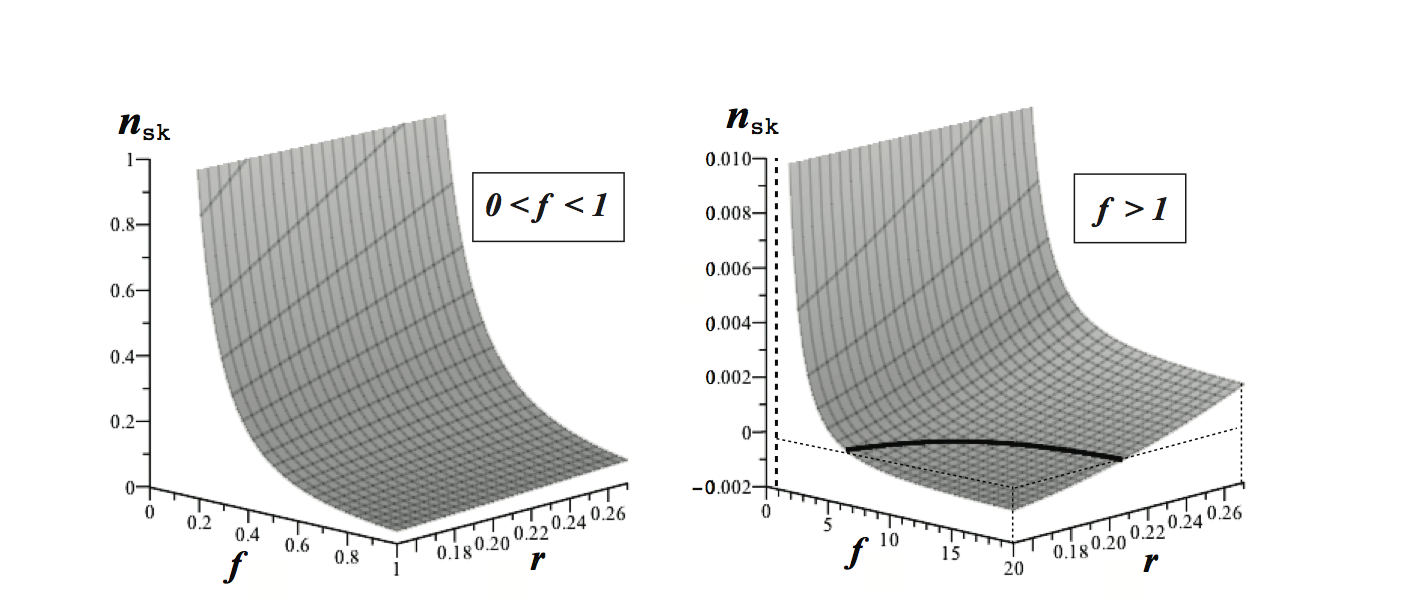}
\caption{\small The running of the spectral index $n_{\mathrm{sk}}
  \equiv\frac{d n_{\mathrm{s}}}{d \ln k}$, Eq.~(\ref{HNInsk}) is here
  shown as a function of the tensor index $r$ and the scale of
  {Goldstone} symmetry breaking $f$ with $\delta _{\mathrm{ns}}\equiv
  1-n_{\mathrm{s}}=1-0.96=0.04$. The left panel shows that
  $n_{\mathrm{sk}}$ is always positive for small values of $f$ (in
  particular for $f < 1$) while, for large $f$ it can become negative
  (right panel). The negative section of the figure is highlighted by
  the level curve at $n_{\mathrm{sk}}=0$. The presence of a large running in hybrid natural inflation could be a possible 
resolution to the apparent tension between high values of $r$ and previous indirect limits based on 
temperature measurements \cite{Ade:2013uln}.}  
\label{f1}
\end{figure}

\begin{figure}[!ht]
\includegraphics[scale=0.32]{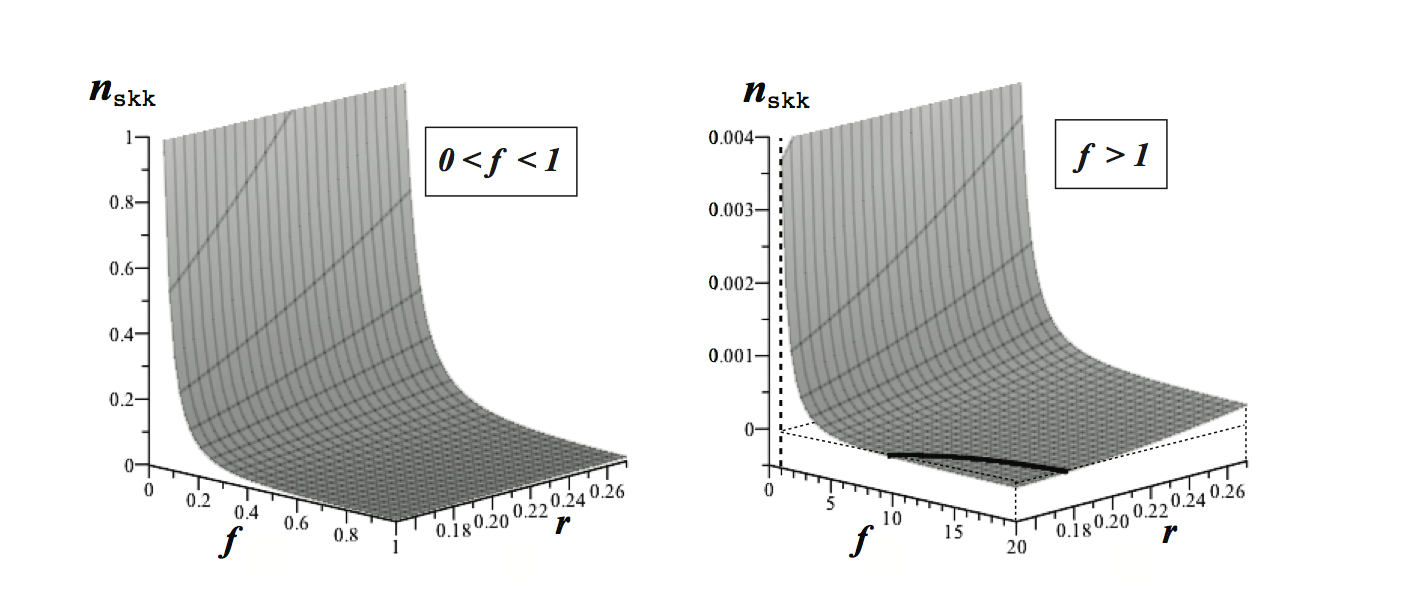}
\caption{\small The running of the running $n_{\mathrm{skk}} \equiv\frac{d^{2}
    n_{\mathrm{s}}}{d \ln k^{2}}$, Eq.~(\ref{HNInskk}) is shown
  as a function of the tensor index $r$ and the symmetry breaking scale $f$ with $\delta _{\mathrm{ns}}\equiv
  1-n_{\mathrm{s}}=1-0.96=0.04$. The left panel shows that
  $n_{\mathrm{skk}}$ is always positive for small values of $f$ while, for large $f$ it can become negative
  (right panel). The negative section of the figure is highlighted by
  the level curve at $n_{\mathrm{skk}}=0$. From Table \ref{table1} we see that the running of the running can be 
high and very close to $10^{-2}.$}
\label{f2}
\end{figure}

Figures~\ref{f1} and \ref{f2} show the running of the spectral index
$n_{\mathrm{sk}}$, Eq.~(\ref{HNInsk}) and the running of the running
$n_{\mathrm{skk}}$, Eq.~(\ref{HNInskk}), respectively as functions of
$f$ and $r$ for the entire range of $r$-values reported by BICEP2. In
both cases the left figures show positive $n_{\mathrm{sk}}$ and
$n_{\mathrm{skk}}$ for small values of $f$ (in particular for $f < 1$) while, for large $f$ both observables can become negative (right). 
The negative section of the figure is highlighted by the level 
curve at $n_{\mathrm{sk}}=0$, $n_{\mathrm{skk}}=0$, respectively.

\begin{figure}[!ht]
\includegraphics[scale=0.45]{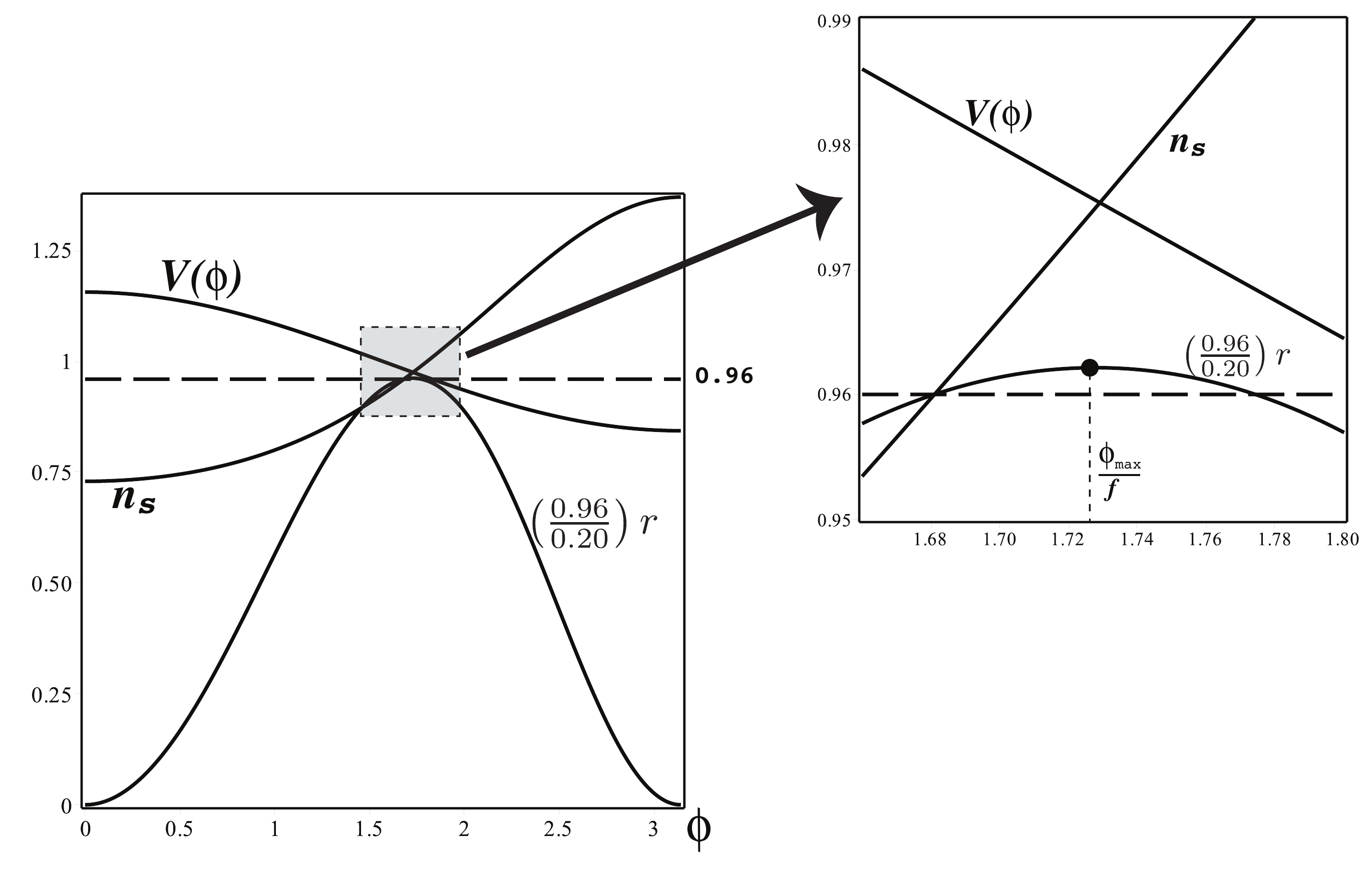}
\caption{\small The plot and inset show the values of the normalised
  potential $V/V_0$, spectral index $n_s$ and a rescaled tensor index
  $r$ in the hybrid natural inflation model.  Values reported by
  BICEP2 are such that  $\phi_{BICEP2} < \phi_{max} $, thus, $r$ grows
  from $\phi_{BICEP2}$ values to the maximum and
  then decreases again to $\phi_{BICEP2}$ values but with the
  spectral index then larger than $0.96$. In the figure we use the set
  of values coming from the Table \ref{table1} where $f=1$, $a=0.156$,
  to obtain $n _{\mathrm{s}}=0.96$, $r=0.20,$ etc. Thus, the evolution
  of $\phi$ during the inflationary period from $\phi_{\mathrm{H}}$ to
  the end of inflation at $\phi_{\mathrm{e}}$  always contains
  $\phi_{max}$ where $r$ reaches a maximum value $r_{max}$ before
  starting to decrease. Consequently far more e-folds are generated
  near the end of inflation (where $r$ is small) rather than close to $\phi_{\mathrm{H}}$.} 
\label{f3}
\end{figure}

Within hybrid natural inflation there is a maximum value that the tensor index can acquire. 
From Eq.~(\ref{HNIeta}) follows that
\begin{equation}
\epsilon =\frac{a^{2}}{2f^{2}}+a^2\eta-\frac{\left( 1-a^{2}\right)f^{2}}{2}\eta^2,
\label{Maxeps}%
\end{equation}%
note that in the case $a=1$, as in natural inflation, this function is
monotonous. In general $\epsilon(\eta)$ presents a maximum where 
$\eta=2\epsilon$, located at
\begin{equation}
\cos \left(\frac{\phi _{max}} {f}\right)=-a .
\label{MaxV}%
\end{equation}%

\noindent Contrary to initial expectations \cite{Hebecker:2014} the
maximum of $r$ (see  Eq.(\ref{HNIeps})) is not located at $
\phi/f=\pi/2$ but at $ \phi_{max}/f=\cos^{-1}(-a)$ and is given by 
\begin{equation} 
r_{max}=8\delta _{\mathrm{ns}}.
\label{Maxr}
\end{equation}
If the maximum occurred in the observable scales, i.e., if $\phi_{max} =
\phi_{\mathrm{H}}$  then $r_{max}=8\delta _{\mathrm{ns}}\approx 0.32.$
This value of $r_{max}$ is larger than the maximum value $r=0.27$
reported by BICEP2. The simple plot of $r$ versus $\phi$
in Figure~\ref{f3} shows that the values reported by BICEP2 are such
that (within hybrid natural inflation) $\phi_{BICEP2} < \phi_{max} $,  
thus, $r$ grows up from $\phi_{BICEP2}$ values to the maximum 
and then decreases again to $\phi_{BICEP2}$ values but with the
spectral index now larger than $0.96$\footnote{The possibility of a
  blue spectrum as a result of the running of the spectral index allows for the 
overproduction of primordial black holes \cite{Kohri:2007qn}. This
aspect will be explored in detail elsewhere}. Thus, during the
evolution of $\phi$ from $\phi_{\mathrm{H}}$ to the end of inflation
at $\phi_{\mathrm{e}}$, the inflationary period always contains
$\phi_{max}$, and thus $r_{max}$. 

Independently of the hybrid natural
inflation case, Eq.~(\ref{Maxr}) can in fact be generalised further to
a critical point; the resulting value $r_{critical}=8\delta
_{\mathrm{ns}}$ is then a universal result \cite{Mariana:2014b}.

\section{Summary and conclusions} \label{Csec}

We find that hybrid natural inflation  can accommodate the recent
results for the tensor index reported by BICEP2  
with values  $\mathcal{O} (M_{Planck})$ of the symmetry breaking scale
$f$ while keeping reasonably small values for the  
other observables. From Table \ref{table1} we see that natural hybrid
inflation predicts a relatively large value for the  
running $n_{\mathrm{sk}}$ which could be observable in the near future. The presence of a large running could be a possible 
resolution to the apparent tension between high values of $r$  and previous indirect limits based on 
temperature measurements. A large running would introduce a scale into the scalar power spectrum to suppress power 
on large angular scales. The running is potentially detectable
by large scale structure through measurements of clustering of high-redshift galaxy surveys 
in combination with CMB data. High-redshift may also be probed with the $21  \rm{cm} $ forest signal observations.
Again from Table \ref{table1} we see that the running of the running can also be large becoming close to $10^{-2}$.

We have also shown that in the context of single field inflation based on the slow-roll paradigm there are equations 
constraining the observables which make no use of any specific potential. 
In particular, failure to satisfy Eq.~(\ref{Slowntk2}) by values obtained for the observables in surveys would indicate a departure from 
the slow-roll approximation itself.

\section{Acknowledgements}

We gratefully acknowledge support from \textit{Programa de Apoyo a Proyectos de Investigaci\'on e Innovaci\'on 
Tecnol\'ogica} (PAPIIT) UNAM, IN103413-3, \textit{Teor\'ias de Kaluza-Klein, inflaci\'on y perturbaciones gravitacionales} 
and IA101414-1, \textit{Fluctuaciones no-lineales en cosmolog\'{\i}a relativista}.  AHA is grateful to the staff of ICF, 
UNAM and UAM-Iztapalapa for hospitality. MCG acknowledges a scholarship from DGAPA-UNAM, IN103413-3 . RAS acknowledges finantial 
support from grant CONACYT 132132. GG, AHA, JCH and RAS thank SNI, 10613 for support.

\end{document}